\documentclass[10pt,twocolumn,superscriptedaddress]{revtex4}
\usepackage[utf8]{inputenc}
\usepackage{hyperref}
\usepackage{epsfig}
\usepackage{graphicx}
\usepackage{amsmath}
\usepackage{amssymb}
\usepackage{ulem}
\usepackage{color}

\graphicspath{{images/}}

\begin{document}

\title{Off-axis compressed holographic microscopy in low light conditions}
\author{Marcio Marim$^{1,2}$, Elsa Angelini$^2$, Jean-Christophe Olivo-Marin$^{1}$ and Michael Atlan$^{3}$}
\affiliation{$^1$Institut Pasteur, Unité d'Analyse d'Images Quantitative, CNRS URA 2582 \\
25-28 rue du Docteur Roux. 75015 Paris. France \\
$^2$Institut Télécom, Télécom ParisTech, CNRS LTCI.
46 rue Barrault. 75013 Paris. France \\
$^3$ CNRS UMR 7587, INSERM U 979, UPMC, UP7, Fondation Pierre-Gilles de Gennes.\\ Institut Langevin. ESPCI ParisTech - 10 rue Vauquelin. 75005 Paris. France \\
}


\begin{abstract}
This article reports a demonstration of off-axis compressed holography in low light level imaging conditions.
An acquisition protocol relying on a single exposure of a randomly undersampled diffraction map of the optical field, recorded in high heterodyne gain regime, is proposed.
The image acquisition scheme is based on compressed sensing, a theory establishing that near-exact recovery of an unknown sparse signal is possible from a small number of non-structured measurements.
Image reconstruction is further enhanced by introducing an off-axis spatial support constraint to the image estimation algorithm.
We report accurate experimental recovering of holographic images of a resolution target in low light conditions with a frame exposure of 5 $\mu$s, scaling down measurements to 9\% of random pixels within the array detector.
OCIS : 070.0070, 180.3170
\end{abstract}

\maketitle

Off-axis holography is well-suited to dim light imaging. Shot-noise sensitivity in high optical gain regime can be achieved with few simple setup conditions \cite{Gross2007}. Holographic measurements are made in dual domains, where each pixel exhibits spatially dispersed (i.e. multiplexed) information from the object. The measurement domain and the image domain are ``incoherent'', which is a requirement for using compressed sensing (CS) sampling protocols \cite{Candes2004b}. In particular, CS approaches using frequency-based measurements can be applied to holography sampling the diffraction field in amplitude and phase. In biological imaging, images are typically compressible or sparse in some domain due to the homogeneity, compactness and regularity of the structures of interest. Such property can be easily formulated as mathematical constraints on specific image features. CS can be viewed as a data acquisition theory for sampling and reconstructing signals with very few measurements \cite{Candes2004b, Candes2006c, Donoho2006}. Instead of sampling the entire data domain and then compress it to take advantage of redundancies, CS enables compressed data acquisition from randomly distributed measurements. Image reconstruction relies on an optimization scheme enforcing some specific sparsity constraints on the image. CS was used recently to improve image reconstruction in holography by increasing the number of voxels one can infer from a single hologram and canceling artifacts \cite{Brady2009, Denis2009, ChoiHorisaki2010}. CS was also used for image retrieval from undersampled measurements in millimeter-wave holography \cite{CullWikner2010} and off-axis frequency-shifting holography \cite{Marim2010a}.\\

In this work, we describe an original acquisition protocol to achieve off-axis compressed holography in low-light conditions, from undersampled measurements. The main result presented in this article is an experimental demonstration of accurate image reconstruction from very few low-light holographic measurements. The acquisition setup consists of a frame exposed with the reference beam alone and subtracted to a frame exposed with light in the object channel, beating against the reference, to yield the holographic signal. This setup prevents any object motion artifact that would potentially occur with phase-shifting methods \cite{Marim2010a}. The CS image reconstruction algorithm relies on a total variation minimization constraint restricted to the actual support of the output image, to enhance image quality.\\

\begin{figure}[]
\centering
\includegraphics[width = 8.0 cm]{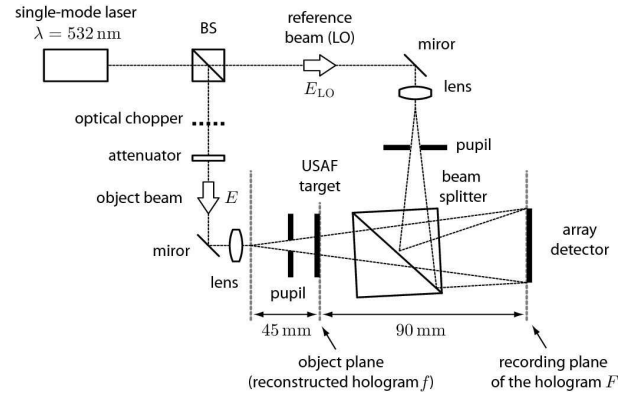}
\caption{Experimental image acquisition setup.}
\label{fig_setup}
\end{figure}

\begin{figure}[]
\begin{center}
\includegraphics[width=8.0 cm]{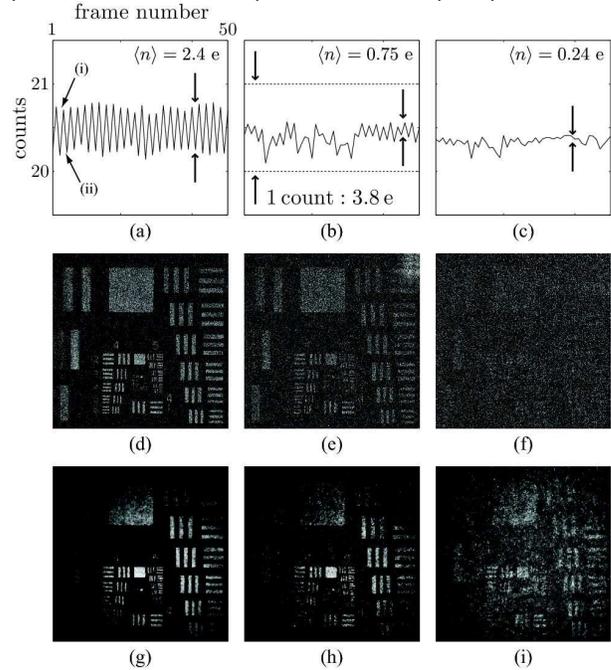}
\end{center}
\caption{Amount of digital counts in the object channel averaged over $N$ pixels, for three different attenuations : $D = 0$ (a), $D = 0.5$ (b), $D = 1$ (c). The LO beam is turned off. The optical field $E$ impinges onto the detector (i) and is blocked (ii) sequentially by the optical chopper, from one frame to the next. The horizontal axis is the frame number, the vertical axis is the average number of counts per pixels. Standard holographic reconstructions at $D = 0$ (d), $D = 0.5$ (e), $D = 1$ (f). CS reconstructions at $D = 0$ with $M/N = 9\%$ (g), at $D = 0.5$ with $M/N = 13\%$ (h), and at $D = 1$ with $M/N = 19\%$ (i).}
\label{fig:light_conditions}
\end{figure}

We consider the holographic detection of an object field $E$ of small amplitude with a reference (or local oscillator) field $E_{\rm LO}$ of much larger amplitude, to seek low-light detection conditions, using the Mach-Zehnder interferometer sketched in fig. \ref{fig_setup}.
%
The main optical radiation comes from a single mode continuous laser at wavelength $\lambda = 532$ nm. Lenses with short focal lengths are used in both channels to create point sources. In the object channel, a negative U.S. Air Force (USAF) resolution target is illuminated in transmission. The amount of optical power in the object channel is tuned with a set of neutral densities. The interference pattern of $E$ beating against $E_{\rm LO}$ corresponds to $I = \left| E + E_{\rm LO} \right|^2$. It is measured within the central region of a Sony ICX 285AL CCD array detector (gain $G_{\rm CCD} = 3.8$ photo-electrons per digital count, $N = N_x \times N_y$ elements, where $N_x = N_y = 1024$, pixel size $d_x = d_y = 6.7 \, \mu \rm m$, quantum efficiency $\sim$ 0.6). The frame rate is set to 12 Hz and the exposure time to 5 $\mu$s. A mechanical chopper is used to switch the object illumination on-and-off from frame to frame. The recorded interference pattern takes the form
\begin{equation}
    I =  |E_{\rm LO}|^2 + |E|^2 + E E_{\rm LO}^* + E^* E_{\rm LO}
\label{Eq_I}
\end{equation}
where $*$ denotes the complex conjugate. In our setup we have $|E|^2 \ll |E_{\rm LO}|^2$. Let's define $n$ and $n_{\rm LO}$, the number of photo-electrons released at each pixel, from light in the object and LO channel respectively, impinging on the detector. The reference beam intensity is adjusted so that the LO shines the detector to half saturation of the pixels' dynamic range, on average. This amounts to $\langle n_{\rm LO} \rangle / G_{\rm CCD} \sim 2000$ digital counts. The brackets $\langle \cdot \rangle$ denote the average over $N$ pixels. Hence $\langle n_{\rm LO} \rangle  = 7.6 \times 10^4$ e (photo-electrons) per pixel. In the object channel, three optical densities $D = 0$, $D = 0.5$, and $D = 1$, are set sequentially to reach very low $\langle n \rangle$ values. The average number of digital counts in 50 consecutive frames recorded in these conditions, while the LO beam is blocked, are reported in figs. \ref{fig:light_conditions}(a)-(c). The detection benefits from a holographic (or heterodyne) gain $G_{\rm H} = \langle |E E_{\rm LO}^*| \rangle / \langle |E|^2 \rangle = (\langle n_{\rm LO} \rangle / \langle n \rangle)^{1/2}$, which ranges from  $G_{\rm H} = 177$ ($D = 0$) to $G_{\rm H} = 563$ ($D = 1$). The spatial support of the signal term $E E_{\rm LO}^*$ is a compact region $\mathcal R$ of $P = 400 \times 400$ pixels. In such high gain regimes, the object field self-beating contribution $|E|^2$, spreading over a region twice as large as $\cal R$ along each spatial dimension, can be neglected in comparison to the magnitude of $E E_{\rm LO}^*$ and $E^* E_{\rm LO}$ in eq. \ref{Eq_I}. In off-axis configuration, the term of interest $E E_{\rm LO}^*$ is also shifted away from $|E|^2$ and $E^* E_{\rm LO}$, which improves the detection sensitivity at the expense of spatial bandwidth. For the current setup, the ratio of available bandwidth between off-axis and on-axis holography is equal to $P/N \sim 15 \%$. To cancel-out the LO flat-field fluctuations within the exposure time, a frame acquired without the object $I_0 = |E_{\rm LO}|^2$ is recorded. The difference of two consecutive frames $I - I_0 \simeq E E_{\rm LO}^* + E^* E_{\rm LO}$ yields a measure of the holographic signal $F = E E_{\rm LO}^*$. $F$ which is proportional to the diffracted complex field $E$, will now be referred to as the optical field itself. Each measurement point on the array detector $F_{p,q}$, where $p = 1, ..., N_x$ and $q = 1, ..., N_y$, corresponds to a point in the Fresnel plane of the object. The optical field $F$, measured in the detection plane yields the field distribution in the object plane $f$ via a discrete Fresnel transform \cite{Schnars1994}
%
\begin{eqnarray}
 \nonumber f_{k,l} = \frac{i}{\lambda \Delta z} {\rm e} ^{ i \pi \lambda \Delta z ( \frac{k^2}{N_x^2 d_x^2} + \frac{l^2}{N_y d_y^2} )}\\
  \times \sum_{p=1}^{N_x} \sum_{q=1}^{N_y} F_{p,q} \, {\rm e} ^{ i \frac{\pi }{\lambda \Delta z} (p^2 d_x^2 + q^2 d_y^2)} \, {\rm e} ^{- 2 i \pi ( \frac{kp}{N_x} + \frac{lq}{N_y} ) }
\label{eq:Fresnel}
\end{eqnarray}
where $i^2 = -1$, and ($k$, $l$), ($p$, $q$) denote pixel indexes. The quadratic phase factor depends on a distance parameter $\Delta z$. Standard holographic reconstruction, as reported in figs. \ref{fig:light_conditions}(d-f), consists in forming the intensity image of the object $g=~|f|^2$ from the measurements of $F$ over the whole detection array, with eq. \ref{eq:Fresnel}.\\

We want to recover $g$ from a small number of measurements $F|_{\Gamma} = \Phi f$ in the detector plane, where $F|_{\Gamma} \subset F$. $\Phi$ is a ($M \times N$) sensing matrix encoding the Fresnel transform (eq. \ref{eq:Fresnel}) and the sampling of a subset $\Gamma$ of $M$ pixels, randomly distributed among the $N$ pixels of the detection array.
We want $M$ to be as small as possible, to benefit from the best compression ratio $M/N$ with respect to non-CS holography, and enhance the throughput savings parameter $1-M/N$. 
For a successful reconstruction, the sensing matrix $\Phi$ must be incoherent with the sparsifying basis $\Psi$ enforced on the reconstructed image \cite{Candes2006c}. This is the case for measurements from the Fresnel transform and a sparsity constraint minimizing the total variation (TV) measure of the reconstructed image $g$. The TV is measured on the gradient map of the image as: $\parallel \nabla g \parallel_{\ell_1} = \sum_{k,l} |\nabla g_{k,l}|$.
Since the target is piecewise constant with sharp edges (such as most microscopy images), its spatial gradient is sparse ${\parallel \nabla g \parallel}_{\ell_0} < M \ll N$. The existence of a sparse representation means that $g$ has at most ${\parallel \nabla g \parallel}_{\ell_0}$ degrees of freedom. For a successful reconstruction we must perform at least $M>{\parallel \nabla g \parallel}_{\ell_0}$ measurements, but much less than $N$.
%
\begin{figure}[]
\centering
\includegraphics[width = 8.0 cm]{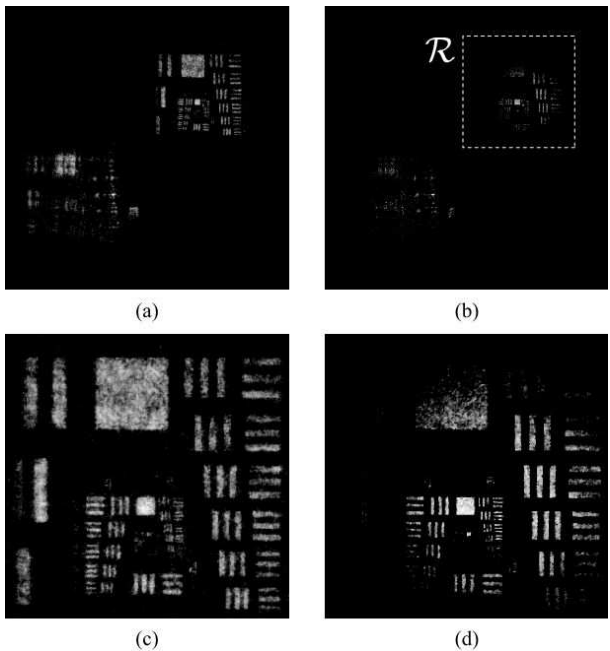}
\caption{Compressed holographic reconstruction of $g$ without support constraint (a). Reconstruction with TV minimization over the region $\cal{R}$ (b). In both cases, $\langle n \rangle = 2.4 \, \rm e$ ($D = 0$) and $M/N = 9\%$. Magnified views over $330 \times 330$ pixels (c,d).}
\label{fig_support}
\end{figure}
Given partial measurements $F|_{\Gamma}$, we seek an estimate $\hat{g}$ with maximum sparsity (i.e. with minimal norm ${\parallel \nabla g \parallel}_{\ell_1}$) whose Fresnel coefficients $\hat{F}|_{\Gamma}$ match the observations $F|_{\Gamma}$ within some error $\delta$. For numerical reasons, the norm ${\parallel \parallel}_{\ell_0}$ was approximated with the norm ${\parallel \parallel}_{\ell_1}$ in the formulation of the reconstruction problem
\begin{equation}
\hat{g} =  \arg \min_{\mathcal{R}} {\parallel \nabla g \parallel}_{\ell_1} \quad \mbox{s.t.} \quad  {\parallel \hat{F}|_\Gamma - F|_\Gamma \parallel}_{\ell_2}\leq \delta
\label{eq:optim_noisy}
\end{equation}
%
$\delta$ is a constraint relaxation parameter introduced to better fit noisy measurements. Contrary to our initial implementation \cite{Marim2010a}, in which the TV measure ${\parallel \nabla \hat{g} \parallel}_{\ell_1}$ was minimized over $N$ reconstructed pixels in the spatial domain, it is now only minimized within the off-axis spatial support $\mathcal{R}$, which bounds are illustrated in Fig. \ref{fig_support}d by the white dashed square. This restriction on the spatial support being constrained leads to a more accurate estimate of $\hat{g}$, actually reducing the number of relevant degrees of freedom to estimate, and hence the number of samples $M$ required. For comparison purposes, CS reconstructions without and with support constraint from the same original frame are reported in fig.\ref{fig_support}. TV minimization over $N$ pixels leads to the hologram magnitude map $\hat{g}$ reported in fig.\ref{fig_support}(a), while the same regularization constraint applied on $\cal R$ leads to the magnitude hologram reported in fig.\ref{fig_support}(b). Magnified views in figs. \ref{fig_support}(c) and \ref{fig_support}(d) show a clear increase in image sharpness with bounded spatial regularization. Noise robustness of compressed holography versus standard holography is illustrated in fig. \ref{fig:light_conditions}, in conditions of low-light illumination of the target. Standard Fresnel reconstruction from $N$ pixels leads to the images reported in figs. \ref{fig:light_conditions} (d)-(f), recorded at $\langle n \rangle = (2.4, 0.75, 0.24)$ for figs.\ref{fig:light_conditions} (d,e,f). CS image reconstructions of $\hat{g}$ with bounded TV regularization from the same data are reported in figs. \ref{fig:light_conditions} (g)-(i). Highly accurate image reconstruction is achieved, at compression rates of 9\% in fig. \ref{fig:light_conditions}(g), 13\% in fig. \ref{fig:light_conditions}(h), and 19\% in fig. \ref{fig:light_conditions}(i), i.e. from much less measurements than needed for Fresnel reconstruction.
%

\bigskip

In conclusion, we have presented a detection scheme for coherent light imaging in low-light conditions successfully employing compressed sensing principles.
It combines a single-shot off-axis holographic scheme, to perform random measurements of an optical field in a diffraction plane, and an iterative image reconstruction enforcing sparsity on a bounded image support.
Compressed off-axis holography is a powerful method to retrieve information from degraded measurements at high noise levels. We demonstrated single-shot imaging in high heterodyne gain regime at 5 $\mu$s frame exposure around one photo-electron per pixel in the object channel. In these conditions, throughput savings from 81\% to 91\% can be reached.\\

This work was funded by Institut Pasteur, DGA, ANR, CNRS, and Fondation Pierre-Gilles de Gennes.\\


\end{document}